\documentclass[aps,twocolumn,floats,prd,nofootinbib,superscriptaddress,10pt]{revtex4-1}

\usepackage{graphicx,amsmath,amssymb,hyperref}
\hypersetup{colorlinks=true, linkcolor=red, filecolor=magenta, urlcolor=blue}
\usepackage{xcolor}

\begin{document}

\title{Closing the window on fuzzy dark matter with the 21cm signal}

\author{Jordan Flitter}
\email{E-mail: jordanf@post.bgu.ac.il}
\author{Ely D.\ Kovetz}
\affiliation{Physics Department, Ben-Gurion University of the Negev, Beer-Sheva 84105, Israel}

\begin{abstract}
Fuzzy dark matter (FDM) is a well motivated candidate for dark matter (DM) as its tiny mass and large de-Broglie wavelength suppress small-scale matter fluctuations, thereby solving some of the small-scale discrepancies in $\Lambda$CDM. Although it has been ruled out as the single component of DM by several observables, there is still a region in the FDM parameter space (the ``FDM window", $10^{-25}\,\mathrm{eV}\lesssim m_\mathrm{FDM}\lesssim10^{-23}\,\mathrm{eV}$) where FDM is allowed to comprise a large portion of the total DM. In this work, for the first time, we study the signature of FDM (comprised of ultra-light axions) in fractions less than unity on the 21cm signal and its detectability by 21cm interferometers such as HERA, taking into account the degeneracy with both astrophysical and cosmological parameters, using  a new pipeline that combines modified versions of the {\tt CAMB} and {\tt 21cmFAST} codes. Our forecasts imply that HERA in its design performance will be sensitive to FDM fractions as small as 1\% in the FDM window, and improve over existing bounds for other masses by up to an order of magnitude.

\end{abstract}

\maketitle

\section{Introduction}

One of the most studied models of DM is ultra-light axions~\cite{Preskill:1982cy, Abbott:1982af, Dine:1982ah, Khlopov:1985jw, Chavanis:2011zi, Kawasaki:2013ae, Hlozek:2014lca, Marsh:2015xka, Hui:2016ltb, Marsh:2016vgj, Desjacques:2017fmf, Hlozek:2017zzf, Fraser:2018acy, Lidz:2018fqo, Poulin:2018dzj, Yang:2019nhz, Ferreira:2020fam, Bauer:2020zsj, Kawasaki:2020tbo, Schutz:2020jox, Sabiu:2021aea, Farren:2021jcd, Marsh:2021lqg, Kawasaki:2021poa, Chadha-Day:2021szb, Hui:2021tkt, Boddy:2022knd, Bernal:2022jap, Laroche:2022pjm}, also known in the literature as fuzzy dark matter (FDM) in the mass range $10^{-27}\,\mathrm{eV}\lesssim m_\mathrm{FDM}\lesssim10^{-20}\,\mathrm{eV}$. FDM became an intriguing candidate for DM as it was shown to ameliorate several problems in cosmology~\cite{Hu:2000ke}, such as the core-cusp problem, the missing satellites problem and the too-big-to-fail problem\footnote{See Ref.~\cite{Bullock:2017xww} and references therein for a review of these problems.}, if the FDM component comprises a meaningful fraction of the total DM. Recently, Ref.~\cite{Blum:2021oxj} proposed that $\mathcal{O}\left(10\%\right)$ of DM in the form of FDM with particle mass in the window  $m_\mathrm{FDM}\approx10^{-25}\,\mathrm{eV}$ could explain the tension between inferences of $H_0$, the current expansion rate of the Universe, which are based on the time-delay in lensed quasar measurements, and those based on CMB observations. Moreover, a fraction $\sim\!10\%$ of FDM was suggested to explain the suppressed amplitude of the matter power spectrum at late times (a.k.a.\ the $\sigma 8$/S8 tension)~\cite{Allali:2021azp,Lague:2021frh,Ye:2021iwa}.

The key physical property of  FDM that impacts the aforementioned observables is that its tiny mass corresponds to an astronomically-large de-Broglie wavelength of $\mathcal{O}\left(\mathrm{1kpc}\right)$. As a result, quantum pressure is exerted on the DM fluid which suppresses small scale fluctuations~\cite{Marsh:2015xka, Marsh:2015daa}. The smaller the FDM mass, the less power to be found on small scales.

\begin{figure*}
\includegraphics[width=0.89\textwidth]{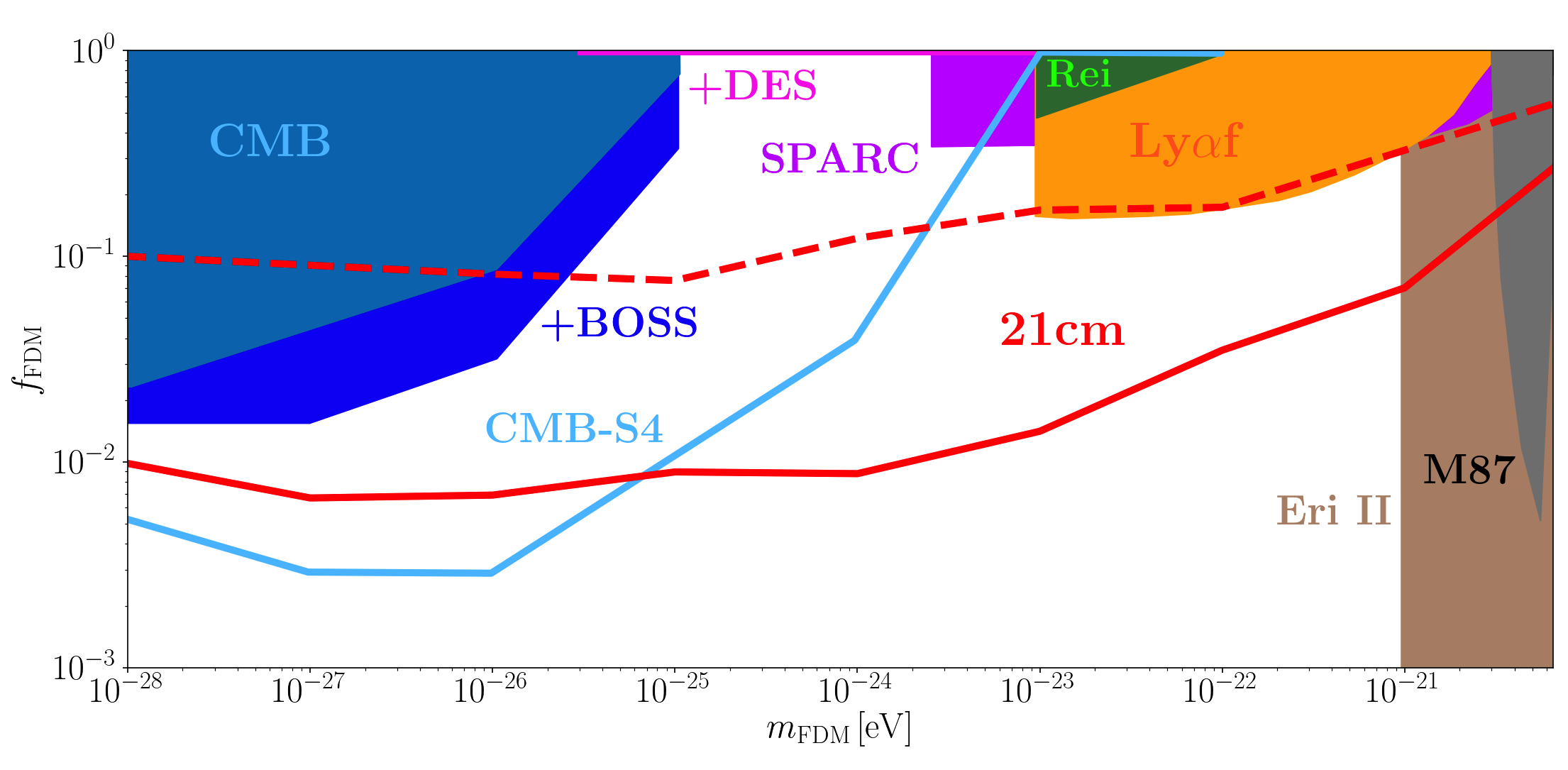}
\vspace{-0.1in}
\caption{Forecasts of the FDM parameter space that can be probed with 21cm interferometers (at a 2-$\sigma$ confidence level). Our 21cm forecasts are shown in \emph{red}, where the solid (dashed) curve corresponds to the lowest limit where full (early) HERA can still probe FDM (assuming a moderate foreground scenario, though results are not much different for pessimistic foregrounds, see the text for details). For comparison, we also present constraints from previous studies. In blue, CMB bounds from Planck~\cite{Hlozek:2014lca, Poulin:2018dzj}, with LSS bounds from galaxy clustering combined with Planck (+BOSS)~\cite{Lague:2021frh}. In orange, bounds from Ly$\alpha$-forest (Ly$\alpha$f)~\cite{Irsic:2017yje, Armengaud:2017nkf, Rogers:2020ltq, Kobayashi:2017jcf}. In green, bounds from the UV luminosity function and optical depth to reionization (Rei.)~\cite{Bozek:2014uqa}. In pink, bounds from galaxy weak lensing combined with Planck (+DES)~\cite{Dentler:2021zij}.  In purple, bounds from galaxy
rotation curves (SPARC)~\cite{Bar:2021kti}. In grey, bounds from the M87 black hole spin, derived from the non-observation of superradiance~\cite{Tamburini:2019vrf, Davoudiasl:2019nlo, Unal:2020jiy}.  In brown, bounds from the half light radius of the central star cluster in the dwarf galaxy Eridanus-II~\cite{Marsh:2018zyw}. In addition, we plot the predicted sensitivity of CMB-S4 to FDM, based on a combined analysis of temperature, polarization and lensing anisotropies~\cite{Hlozek:2016lzm}.}
\label{constraints_plot}
\end{figure*}

The small-scale signature of FDM has been well studied in the literature. Refs.~\cite{Hlozek:2014lca, Lague:2021frh} obtained constraints on the FDM fraction from CMB and large scale structure (LSS) measurements. They concluded that if FDM exists in the mass range $10^{-32}\,\mathrm{eV}\lesssim m_\mathrm{FDM}\lesssim10^{-26}\,\mathrm{eV}$ it can only comprise a few percent of the total DM in the Universe. For higher values of $m_\mathrm{FDM}$, the bound becomes weaker because the FDM power suppression becomes less efficient, while for smaller values of $m_\mathrm{FDM}$ the FDM behaves as dark energy (DE) and is strongly degenerate with $\Omega_\Lambda$. Another constraint on FDM was obtained from Ly$\alpha$-forest data, indicating that DM cannot be fully described by FDM with mass of $10^{-21}\,\mathrm{eV}\lesssim m_\mathrm{FDM}\lesssim2\times10^{-20}\,\mathrm{eV}$~\cite{Irsic:2017yje, Armengaud:2017nkf, Rogers:2020ltq}, while FDM with mass of $10^{-22}\,\mathrm{eV}\lesssim m_\mathrm{FDM}\lesssim10^{-21}\,\mathrm{eV}$ must not comprise more than $\mathcal O\left(10\%\right)$ of the total DM~\cite{Kobayashi:2017jcf}. Furthermore, the reionization history of our Universe, inferred from the UV-luminosity function and optical depth to reionization, excludes DM models where the FDM fraction is 50\% or higher for $m_\mathrm{FDM}$ down to $10^{-23}\,\mathrm{eV}$~\cite{Bozek:2014uqa}. Finally, data from the Dark Energy Survey year 1 (DES-Y1)~\cite{DES:2018zzu} was used to search for shear-correlation suppression caused by FDM, ruling out the existence of FDM in the mass range $10^{-25}\,\mathrm{eV}\lesssim m_\mathrm{FDM}\lesssim10^{-23}\,\mathrm{eV}$, if DM is entirely made by FDM~\cite{Dentler:2021zij}. However, current constraints suggest that it is possible for FDM to exist in large portions in the region $10^{-25}\,\mathrm{eV}\lesssim m_\mathrm{FDM}\lesssim10^{-23}\,\mathrm{eV}$ (referred to below as the {\it FDM window}), as can be seen in Fig.~\ref{constraints_plot}.

In this work we explore in detail the signature of FDM on the 21cm signal. Previous works \cite{Jones:2021mrs, Hotinli:2021vxg, Nebrin:2018vqt, Sarkar:2022dvl} have demonstrated the delaying effect of FDM on the 21cm signal due to impeding  structure formation, but have all assumed an FDM fraction of unity, an assumption that is relaxed in this work. Unlike Refs.~\cite{Marsh:2015daa,Hotinli:2021vxg}, which evolved the FDM evolution equations in the fluid approximation starting from matter-radiation equality (MRE) or even at recombination, we start evolving the FDM field equation before MRE to yield consistent initial conditions for the 21cm signal simulation. We apply our analysis to the Hydrogen Epoch of Reionization Array (HERA)~\cite{DeBoer:2016tnn} (for an analytic estimate of forecasts for much longer down the road, with the Square Kilometer Array, see Ref~\cite{Giri:2022nxq}). Our results, presented by the red curves in Fig~\ref{constraints_plot}, suggest that HERA will be sensitive to FDM fractions of $f_\mathrm{FDM}\gtrsim\mathcal O\left(1\%\right)$ for a wide range of FDM masses, $10^{-28}\,\mathrm{eV}\lesssim m_\mathrm{FDM}\lesssim10^{-23}\,\mathrm{eV}$. We note that our forecasts are stronger than most current bounds on FDM, are competitive with those of the CMB-S4 experiment where they overlap and extend well beyond its reach. This demonstrates the great potential of 21cm interferometers as a means to probe various models of DM.

The remaining parts of this paper are organized as follows. Section~\ref{sec: Formalism} provides a brief introduction to the 21cm signal and the underlying physics of FDM. In Section~\ref{sec: Simulation} we describe how we simulate the 21cm signal in an FDM universe, and show the impact of different FDM parameters on the 21cm signal as well as their degeneracies. We discuss our forecasts in Section~\ref{sec: Forecasts} and conclude in Section~\ref{sec: Conclusions}.

\section{Formalism}\label{sec: Formalism}
The origin of the 21cm signal is the 21cm-wavelength photon that is emitted  whenever a hydrogen atom undergoes the hyperfine transition from the triplet state (where the spins of the electron and proton are aligned) to the singlet state (where the spins are anti-aligned)~\cite{Madau:1996cs,Furlanetto:2006jb,Pritchard:2011xb}. Between recombination and reionization, the Universe was permeated with neutral hydrogen atoms that could get excited to the triplet state through absorption of CMB photons, collisions between the gas particles, and Ly$\alpha$ pumping (Wouthuysen-Field effect~\cite{1952AJ.....57R..31W,1958PIRE...46..240F}) as a result of the radiation from the first generations of stars. 

The cosmological 21cm signal is manifested as a brightness temperature,
\begin{equation}
T_{21}=\frac{T_s-T_\mathrm{rad}}{1+z}\left(1-\mathrm{e}^{-\tau_{21}}\right),
\end{equation}
where $T_\mathrm{rad}$ is usually assumed to be the CMB temperature, $T_\mathrm{CMB}\propto\left(1+z\right)$, and $\tau_{21}$ is the 21cm optical depth. The quantity $T_s$ is known as the spin temperature, and is defined by the ratio between the number densities of hydrogen atoms in the triplet and singlet states, $n_1/n_0=3\exp\left(-\frac{h\nu_{21}}{k_BT_s}\right)$, where $h$  and $k_B$ are the Planck and Boltzmann constants, and $\nu_{21}\approx1420\,\mathrm{MHz}$ is the frequency of the 21cm photon. In thermal equilibrium, the spin temperature is given by~\cite{Sarkar:2022dvl}
\begin{equation}
T_s^{-1}=\frac{x_\mathrm{rad}T_\mathrm{rad}^{-1}+x_cT_k^{-1}+\tilde x_\alpha T_c^{-1}}{x_\mathrm{rad}+x_c+\tilde x_\alpha},
\end{equation}
where $T_k$ is the gas kinetic temperature, $T_c\approx T_k$ is the effective color temperature of the Ly$\alpha$ photons, and $x_\mathrm{rad}$, $x_c$ and $\tilde x_\alpha$ are the radiation, collision, and Ly$\alpha$ coupling coefficients, respectively.

Because $\tilde x_\alpha$ is proportional to the Ly$\alpha$ flux $J_\alpha$~\cite{Hirata:2005mz}, and $J_\alpha$ is in turn proportional to the integrated value of the star formation rate (SFR), the onset of cosmic dawn can be identified as the moment when $\tilde x_\alpha$ is dominant. At that time, the spin temperature closely follows $T_k$, and because the gas cools adiabatically, i.e.\ faster than the CMB, we expect to see an absorption signal in the brightness temperature. For $\Lambda$CDM, the location of the minimum of $T_{21}$ strongly depends on the assumed astrophysical model, but is expected to be found at $10\lesssim z\lesssim20$.

However, the minimum of the global 21cm signal can shift towards smaller redshifts if the Universe contains large portions of FDM. The FDM particle is the excited state of the scalar field $\phi$ which obeys the Klein-Gordon equation, $\left(\square-m_\mathrm{FDM}^2\right)\phi=0$~\cite{Hu:2000ke, Hlozek:2014lca, Marsh:2015xka, Marsh:2015daa, Hui:2021tkt}. Initially, the field slowly rolls down its potential when $m_\mathrm{FDM}\ll H$ and behaves as DE with equation of state $w=-1$, where $H$ is the Hubble parameter. However, as $m_\mathrm{FDM}\sim3H$, the field begins to oscillate and its energy density dilutes as for matter, $\rho_\mathrm{FDM}\propto a^{-3}$, where $a$ is the scale factor of the Universe~\cite{Marsh:2015xka, Hlozek:2014lca}. If the mass of the field is $m_\mathrm{FDM}\gtrsim10^{-28}\,\mathrm{eV}$ (which we assume throughout this work), the oscillation phase of the field starts well before MRE, and the field can be characterized as matter throughout its evolution. In that case, the impact of the FDM field on the Boltzmann equation for the dark matter fluid comes in the form of new terms that are proportional to $c_\mathrm{s,a}^2$, where $c_\mathrm{s,a}$ is the FDM (or axion) sound speed, and is given by\footnote{In natural units where $c=\hbar=1$.}~\cite{Hlozek:2014lca}
\begin{equation}
c_\mathrm{s,a}^2=\frac{k^2/\left(4m_\mathrm{FDM}^2a^2\right)}{1+k^2/\left(4m_\mathrm{FDM}^2a^2\right)}.
\end{equation}

Thus, while for large scales ($k\ll2m_\mathrm{FDM}a$) the evolution of the matter-density fluctuations remains the same as in $\Lambda$CDM, small-scale fluctuations ($k\gg2m_\mathrm{FDM}a$) in an FDM universe are washed out. Ref.~\cite{Hu:2000ke} found that the scale at which the matter power spectrum drops by a factor of 2 is $k_{1/2}\approx4.5\left(m_\mathrm{FDM}/10^{-22}\,\mathrm{eV}\right)^{4/9}\,\mathrm{Mpc}^{-1}$, so smaller $m_\mathrm{FDM}$ implies less power, and therefore structure formation is impeded along with the SFR, consequently leading to a shift of the minimum of the 21cm signal towards smaller redshifts. This behavior was demonstrated in Refs.~\cite{Jones:2021mrs, Hotinli:2021vxg, Nebrin:2018vqt, Sarkar:2022dvl} where FDM was assumed to comprise the entirety of DM in the Universe. Yet, if FDM makes up only a small fraction of the total DM, its signature in the 21cm cosmic dawn signal might be completely lost.

\section{Simulation}\label{sec: Simulation}
We simulate the effect of the FDM mass, $m_\mathrm{FDM}$, and fraction, $f_\mathrm{FDM}\equiv\Omega_\mathrm{FDM}/\Omega_\mathrm{DM}$, on the matter transfer function with {\tt AxionCAMB}\footnote{\href{https://github.com/dgrin1/axionCAMB}{github.com/dgrin1/axionCAMB}}~\cite{Hlozek:2014lca}, a modified version of {\tt CAMB}~\cite{Lewis:1999bs} which takes into account the impact of the FDM on the Boltzmann equations even before MRE. 
We then study the 21cm signal in an FDM universe by feeding the matter transfer function obtained with {\tt AxionCAMB} into {\tt 21cmFast}\footnote{\href{https://github.com/21cmfast/21cmFAST}{github.com/21cmfast/21cmFAST}} version 3.1.3~\cite{Munoz:2021psm}. In this version, the Lyman-Werner (LW) feedback~\cite{Haiman:1996rc, Bromm:2003vv} as well as the relative velocity between baryons and DM ($v_\mathrm{cb}$)~\cite{Tseliakhovich:2010bj, Ali-Haimoud:2013hpa} are taken into account in each voxel. As these effects suppress structure formation on small scales, they delay cosmic dawn, much like FDM. In addition, version 3.1.3 of {\tt 21cmFast} separates the pop-II and pop-III stars into atomic and molecular cooling galaxies (MCGs), respectively. We have also modified {\tt 21cmFast} and incorporated the Ly$\alpha$ heating mechanism~\cite{Chen:2003gc, Reis:2021nqf, Sarkar:2022dvl}. 

Our fiducial values for the cosmological parameters were taken from Planck 2018~\cite{Planck:2018vyg} while the fiducial values for the astrophysical parameters were taken from Table 1 (EOS2021) of Ref.~\cite{Munoz:2021psm}, except we set $L_X^\mathrm{(II)}=L_X^\mathrm{(III)}=39$ to compensate for the extra heating caused by Ly$\alpha$ heating (see description of the free parameters in our analysis below Eq.~\ref{Eq: free parameters}). In our simulations, we use a box size of $200\,\mathrm{Mpc}$ and resolution of $1/3\,\mathrm{Mpc}$. This is a fairly small box, but its high resolution allowed us to better probe the fluctuations on small scales, which are the ones most affected by FDM. Nevertheless, we have confirmed that our results are insensitive to the simulation box size.

\subsection{The 21cm global signal}
\begin{figure}
\includegraphics[width=\columnwidth]{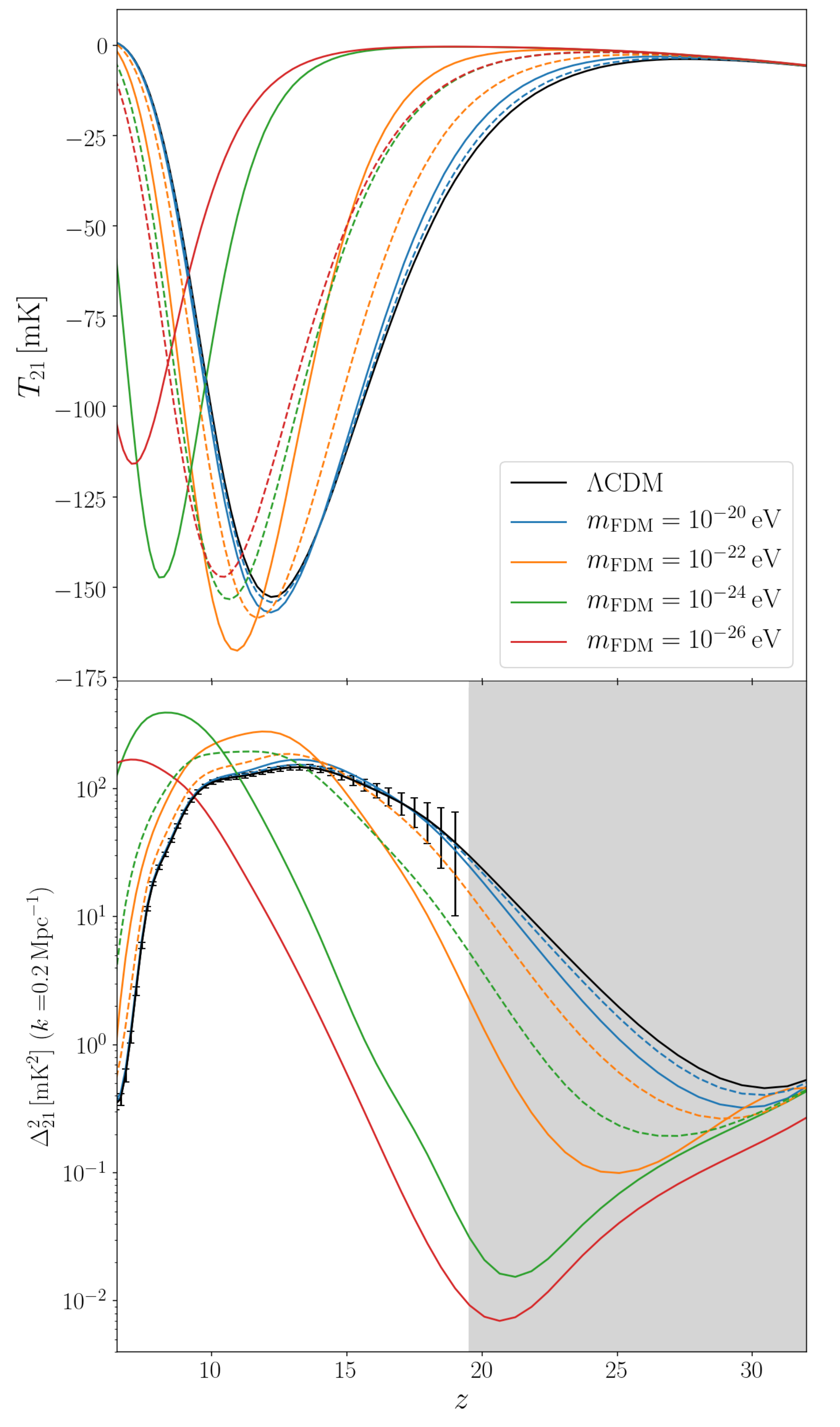}
\vspace{-0.25in}
\caption{\emph{Top panel:} the 21cm global signal for several FDM masses. Solid (dashed) curves correspond to fractions $f_\mathrm{FDM}\!=\!10\%$ ($3\%$). \emph{Bottom panel:} the 21cm power spectrum at $k=0.2\mathrm{Mpc^{-1}}$. The error bars correspond to the noise of full HERA, calculated with {\tt 21cmSense} for moderate foregrounds scenario (see Section \ref{sec: Degeneracies}), while the shaded region corresponds to redshifts where the noise surpasses the signal, $\delta\Delta^2_{21}>\Delta^2_{21}$. For this foregrounds scenario, HERA can detect the signal with a finite uncertainty (i.e. beyond the horizon wedge) at scales between $0.1\lesssim k\lesssim0.6\,\mathrm{Mpc}^{-1}$, where the noise mildly increases as $k$ increases.}
\label{global_signal}
\end{figure}

We demonstrate in Fig.~\ref{global_signal} (top panel) how the 21cm global signal is affected by FDM with varying $m_\mathrm{FDM}$ and $f_\mathrm{FDM}=3\%, 10\%$. As expected, the absorption signal shifts to smaller redshifts as we decrease the FDM particle mass. Evidently, the delaying effect of FDM is more pronounced as its fraction is greater, but the dependence of the brightness temperature on $f_\mathrm{FDM}$ only becomes evident at small $m_\mathrm{FDM}$. Naturally, the information encoded in the power spectrum dwarfs that in the global signal and allows to break degeneracies with competing effects~\cite{Sarkar:2022dvl}.

\subsection{The 21cm power spectrum}
Our goal is to examine what is the lowest $f_\mathrm{FDM}$ that 21cm interferometers such as HERA will be able to probe. In order to answer that question, we use the {\tt powerbox}\footnote{\href{https://github.com/steven-murray/powerbox}{github.com/steven-murray/powerbox}} module~\cite{2018JOSS....3..850M} to compute the 21cm power spectrum from the lightcone box produced by {\tt 21cmFast},
\begin{equation}\label{Eq: 21cm power}
\Delta_{21}^2\left(k,z\right)=\frac{k^3P_{21}\left(k,z\right)}{2\pi^2},
\end{equation}
where $P_{21}\left(k,z\right)=\langle \tilde T_{21}(\vec k,z)\tilde T^*_{21}(\vec k,z)\rangle$, and $\tilde T_{21}(\vec k,z)$ is the Fourier transform of $T_{21}\left(\vec x,z\right)-\langle T_{21}\left(z\right)\rangle$.

In the bottom of Fig.~\ref{global_signal} we show $\Delta_{21}^2\left(z\right)$ for $k=0.2\,\mathrm{Mpc}^{-1}$. As a zero order approximation, we see from Eq.~\eqref{Eq: 21cm power} that $\Delta^2_{21}\left(z\right)\propto\langle T_{21}\left(z\right)\rangle^2$. In other words, we expect the 21cm power spectrum to follow the global signal. This feature is evident in Fig.~\ref{global_signal}, and in particular we can see that the power spectrum is also delayed by FDM. However, there are some exceptions --- for example, the amplitude of the global signal for $m_\mathrm{FDM}=10^{-22}\,\mathrm{eV}$ is greater than the amplitude for $m_\mathrm{FDM}=10^{-24}\,\mathrm{eV}$, but the power spectrum of the former reaches to smaller values compared to the latter. Moreover, on large scales $\Delta^2_{21}\left(z\right)$ exhibits additional features~\cite{Sarkar:2022dvl}. Regardless, to understand our forecasts the approximation  $\Delta^2_{21}\left(z\right)\propto\langle T_{21}\left(z\right)\rangle^2$ is still useful.

\subsection{Degeneracies}\label{sec: Degeneracies}
To calculate the uncertainty in the measurement of $\Delta_{21}^2\left(k,z\right)$ (which we denote by $\delta\Delta_{21}^2\left(k,z\right)$), we use {\tt 21cmSense}\footnote{\href{https://github.com/steven-murray/21cmSense}{github.com/steven-murray/21cmSense}}~\cite{Pober:2012zz, Pober:2013jna} and simulate HERA at its design performance (\emph{full} HERA). We use the built-in configuration of a hexagonal array with 11 antennae at its base (311 antennae overall), each of which has a diameter of $14\,\mathrm{m}$, and we assume a receiver temperature of $T_\mathrm{rec}=100\,\mathrm{K}$. The frequency range for HERA is taken to be $50-225\,\mathrm{MHz}$ and we divide the corresponding redshift range into 15 bins. We assume that HERA operates six hours per night for three years with bandwidth $8\,\mathrm{MHz}$, and we assume the sky temperature obeys $T_\mathrm{sky}\left(\nu\right)=60\,\mathrm{K}\left(\nu/300\,\mathrm{MHz}\right)^{-2.55}$. We also assume in our analysis the moderate (pessimistic) foregrounds scenario in which the wedge is assumed to extend to $\Delta k_{||}=0.1h\mathrm{Mpc}^{-1}$ beyond the horizon wedge limit, and all baselines are added coherently (incoherently).

We perform a Fisher analysis and calculate the Fisher matrix according to~\cite{Jungman:1995bz, Bassett:2009tw, Jimenez:2013mga, Munoz:2019hjh}
\begin{equation}
F_{\alpha,\beta}=\sum_{k,z}\frac{\partial \Delta_{21}^2\left(k,z\right)}{\partial\alpha}\frac{\partial \Delta_{21}^2\left(k,z\right)}{\partial\beta}\frac{1}{\left[\delta\Delta_{21}^2\left(k,z\right)\right]^2},
\end{equation}
where $\alpha$ and $\beta$ represent the varied parameters of the model. In our analysis, they are 
\begin{flalign}\label{Eq: free parameters}
\nonumber&\left(\alpha,\beta\right)\in\{h,\Omega_m,\Omega_b,A_s,\log_{10}m_\mathrm{FDM},\log_{10}f_\mathrm{FDM},&
\\&\hspace{15mm}\log_{10}L_X^{(\mathrm{II})},\log_{10}f_*^{(\mathrm{II})},\log_{10}f_\mathrm{esc}^{(\mathrm{II})},A_\mathrm{LW},A_{v_\mathrm{cb}}\},& 
\end{flalign}
where $h$ is Hubble's constant, $\Omega_m$ ($\Omega_b$) is the matter (baryons) density parameter, $A_s$ is the primordial curvature fluctuation amplitude ,$L_X^{(\mathrm{II})}$ is the X-ray luminosity of pop-II stars (normalized by the SFR, in units of $\mathrm{erg}\,\mathrm{s}^{-1}\,M_\odot^{-1}\,\mathrm{yr}$), $f_*^{(\mathrm{II})}$ is the pop-II star formation efficiency, $f_\mathrm{esc}^{(\mathrm{II})}$ is the fraction of photons that were produced by pop-II stars and managed to escape the host galaxy and ionize the inter-galactic medium, and $A_\mathrm{LW}$ ($A_{v_\mathrm{cb}}$) is the amplitude of the LW feedback ($v_\mathrm{cb}$) on the minimum MCG mass that can still host stars\footnote{See more details on the exact definitions of the astrophysical parameters of Eq.~\eqref{Eq: free parameters} in Ref.~\cite{Munoz:2021psm}.}. We note that $m_\mathrm{FDM}$ in Eq.~\eqref{Eq: free parameters} is given in units of eV. Plus, we note that whenever an astrophysical pop-II parameter is varied in our analysis, the corresponding pop-III parameter is varied as well.

\begin{figure}
\includegraphics[width=\columnwidth]{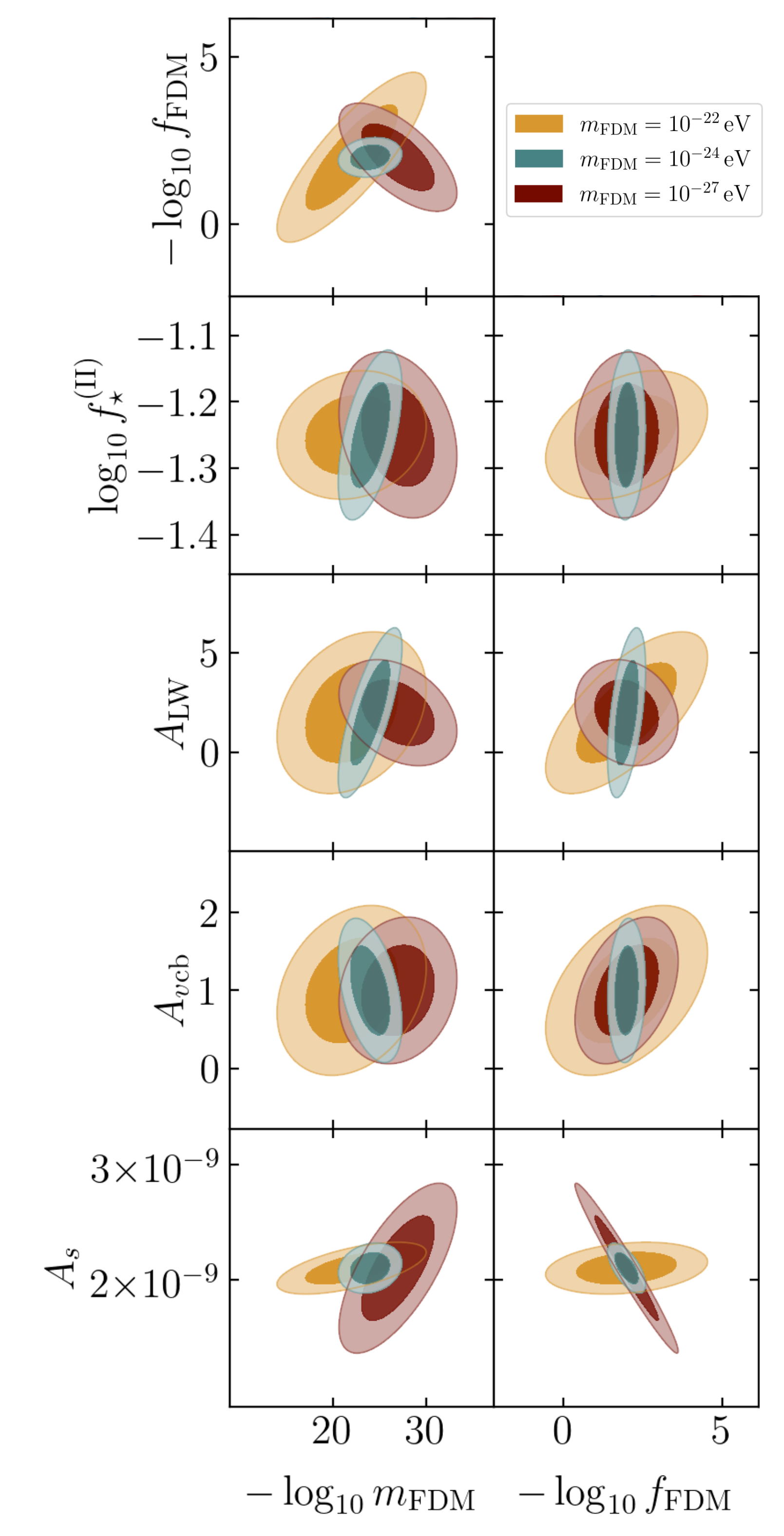}
\vspace{-0.15in}
\caption{Degeneracies of the FDM parameters, for $f_\mathrm{FDM}=1\%$. The ellipses correspond to 1-$\sigma$ and 2-$\sigma$ confidence levels. Here, we take no priors and marginalize over the rest of the parameters in Eq.~\eqref{Eq: free parameters}.}
\vspace{-0.15in}
\label{Fisher ellipses}
\end{figure}

The sensitivity of HERA to FDM depends on the degeneracies between the FDM parameters and other cosmological/astrophysical parameters that shape the 21cm signal. Once we have the Fisher matrix at our disposal, we can study these degeneracies, as we do in Fig.~\ref{Fisher ellipses} (to clearly see the degeneracies, we do not impose any priors at this stage). We can clearly see how the degeneracies change as $m_\mathrm{FDM}$ is varied. At large masses ($m_\mathrm{FDM}\gtrsim10^{-22}\,\mathrm{eV}$) $m_\mathrm{FDM}$ and $f_\mathrm{FDM}$ are positively correlated, while at small masses ($m_\mathrm{FDM}\lesssim10^{-27}\,\mathrm{eV}$) they exhibit negative correlation. This feature is due to the unique properties of the FDM at the extreme sides of the mass spectrum. At the large mass regime, smaller $m_\mathrm{FDM}$ tends to delay cosmic dawn. On the other end, if $m_\mathrm{FDM}\lesssim10^{-27}\,\mathrm{eV}$, the field $\phi$ is on the verge of being DE-like~\cite{Hlozek:2014lca} and contributes to the DE energy budget, so reducing $m_\mathrm{FDM}$ while lowering the value of $\Omega_\Lambda$ to maintain the flatness of the Universe) yields greater power at all scales, shifting  cosmic dawn to earlier redshifts. However, in both $m_\mathrm{FDM}$ regimes, smaller $f_\mathrm{FDM}$ impacts the cosmic dawn in the same manner, by shifting it to higher redshifts. Another strong degeneracy between $A_s$ and $f_\mathrm{FDM}$ can be seen in the small $m_\mathrm{FDM}$ regime, where increasing $f_\mathrm{FDM}$ is equivalent to reducing $A_s$. Although $A_\mathrm{LW}$ and $A_{v_\mathrm{cb}}$ impose a delaying effect on the 21cm signal~\cite{Sarkar:2022dvl, Munoz:2019rhi}, similarly to FDM, we find that their degeneracy with the FDM parameters is not significant.

\section{Forecasts}\label{sec: Forecasts}
In order to derive the sensitivity of HERA  to FDM, we now add priors on the cosmological parameters from Planck 2018~\cite{Planck:2018vyg}, take the inverse matrix, and find the 1-$\sigma$ uncertainty of $f_\mathrm{FDM}$, $\sigma\left(f_\mathrm{FDM}\right)$, by marginalizing over the rest of the parameters of Eq.~\eqref{Eq: free parameters}. For each $m_\mathrm{FDM}$ we simulate, we repeat the above process for several values of $f_\mathrm{FDM}$, and solve for $f_\mathrm{FDM}=2\sigma\left(f_\mathrm{FDM}\right)$ to find the 2-$\sigma$ forecasts. We show our results in Fig.~\ref{constraints_plot} with respect to the moderate foreground scenario (solid curve).

As expected, for large $m_\mathrm{FDM}$ the sensitivity of HERA monotonically increases as the mass of the FDM decreases. This is because small $m_\mathrm{FDM}$ implies greater suppression in the matter power spectrum, which causes additional delay in the global signal as we demonstrated in Fig.~\ref{global_signal}. Since the observable, $\Delta_{21}^2$, is roughly proportional to the square of the global signal, the 21cm power spectrum exhibits the same delaying feature. The suppression/delaying effect of FDM becomes more pronounced as its mass further decreases, rendering it more detectable. Moreover, as the 21cm signal gets pushed to smaller redshifts (by decreasing $m_\mathrm{FDM}$ accordingly), the sensitivity of HERA improves. For $m_\mathrm{FDM}=10^{-24}\,\mathrm{eV}$, HERA is still sensitive to FDM with a fraction of $f_\mathrm{FDM}\sim\mathcal O\left(1\%\right)$. For such a small fraction, even when $m_\mathrm{FDM}$ keeps decreasing, the impact of the FDM on the 21cm signal saturates, which is the reason for the plateau in Fig.~\ref{constraints_plot}. At $m_\mathrm{FDM}\lesssim10^{-27}\,\mathrm{eV}$, as was discussed earlier, this is the regime where the field $\phi$ is DE-like and an upward trend in our forecasts can be seen. In this regime, for a given fraction, as $m_\mathrm{FDM}$ decreases the field $\phi$ behaves more similarly to DE, and thus for a given set of cosmological parameters the matter power spectrum approaches the $\Lambda$CDM prediction, making FDM (or more precisely at this regime, the axions) less detectable.

\section{Conclusions}\label{sec: Conclusions}
In this work we have simulated the 21cm signal in universes where FDM comprises only a fraction of the total DM. Motivated by the large delaying effect that FDM imposes on the 21cm signal, even at small fractions, we simulated the sensitivity of HERA to FDM.
Our results indicate that HERA can yield constraints on FDM which are nearly twice tighter than current CMB +LSS bounds (for $m_\mathrm{FDM}\lesssim10^{-26}\,\mathrm{eV}$) and are competitive with CMB-S4 forecasts~\cite{Hlozek:2016lzm}, and constraints that are stronger by an order of magnitude than current Ly$\alpha$-forest bounds (for $m_\mathrm{FDM}\gtrsim10^{-23}\,\mathrm{eV}$). 
In the poorly constrained {\it FDM window}, $10^{-25}\,\mathrm{eV}\lesssim m_\mathrm{FDM}\lesssim10^{-23}\,\mathrm{eV}$, we forecast that HERA will be sensitive to  $f_{\rm FDM}\gtrsim\mathcal O\left(1\%\right)$. We note that our forecasts are robust to changes in the fiducial values of the astrophysical parameters\footnote{We have recalculated the forecasts for a fiducial case where $L_X^{(\mathrm{II})}$ and $f_*^{(\mathrm{II})}$ are stronger by an order of magnitude. For $\Lambda$CDM, this leads to an absorption signal at $z\approx17$ and a later emission feature which is absent in Fig.~\ref{global_signal}. For the masses we have tested, the deviation in the forecasts was no more than a few percent.}, or considering a pessimistic foreground scenario.

With these results, we could not resist the temptation of deriving the potential constraining power of HERA in its early stage (\emph{early} HERA). Following Ref.~\cite{HERA:2021noe}, we modelled early HERA as a hexagon core with seven antennae at its base (127 antennae overall) and  assumed a total operation time of one year. We also limited the frequency range to $100\!-\!200\mathrm{MHz}$ and used only four redshift bins, centered at $z=7.5,8.3,9.3,10.1$. All the other specifications remain the same. With these settings, we recalculated the forecasts and plotted them as a dashed curve in Fig.~\ref{constraints_plot}. The constraining power of early HERA is degraded by an order of magnitude compared to the constraining power of full HERA, but the results are still impressive nonetheless. Provided there are no significant systematics~\cite{Hills:2018vyr, Bradley:2018eev, Kern:2019ytc, Liu:2019awk}, we find that early HERA is able to detect FDM in a wide range of masses, including the {\it FDM window}, if $f_\mathrm{FDM}\gtrsim\mathcal O\left(10\%\right)$.

In our analysis, we have adopted the phenomenological astrophysical model of Refs.~\cite{Munoz:2021psm, Park:2018ljd}. According to this model, the star formation efficiencies  are redshift independent and assumed to be a power-law in halo mass. This is the standard model of {\tt 21cmFast} version 3.1.3, although photo-heating and reionization feedbacks at $z\lesssim 7$ seem to lower both the amplitude of the star formation efficiency and the power-law index at small halos~\cite{Wu:2019wnd, Ocvirk:2018pqh}, thus rendering the SFR to have more complicated time dependence than implemented in {\tt 21cmFast}. We leave the study of advanced time dependent star formation efficiency models (along with studying degeneracies of the FDM parameters with additional astrophysical parameters) for future work.

Our main conclusion is that in the near future HERA could discover evidence for FDM in the form of ultralight axions with unrivaled sensitivity in the {\it FDM window}. This demonstrates the crucial role that 21cm interferometers will have in unveiling the fundamental properties of some (if not all) the dark matter in our Universe.

\begin{acknowledgments}

We thank Debanjan Sarkar for reviewing our implementation of the Ly$\alpha$ heating in {\tt{21cmFast}} version 3.1.3. We would like to thank David Marsh, Selim Hotinli and  Edoardo Vitagliano for useful discussions. We would like also to thank the anonymous referees for their comments that helped to improve the quality of the paper. JF is supported by the Zin fellowship awarded by the BGU Kreitmann School. EDK acknowledges support from an Azrieli faculty fellowship.
\end{acknowledgments}

\end{document}